\documentclass[epjST]{svjour}

\usepackage[utf8]{inputenc}
\usepackage[T1]{fontenc}
\usepackage{graphics}
\usepackage[square, numbers]{natbib}
\usepackage{xcolor}
\usepackage{hyperref}
\usepackage{hyphenat}
\usepackage{amsmath,amsfonts,amssymb,array}
\usepackage{mathabx}
\usepackage{lscape}

\begin{document}

\title{Modelling nonlinear dynamics of interacting tipping elements on complex networks: the \emph{PyCascades} package}
\author{Nico Wunderling\inst{1,2,3,}\footnote{Correspondences should be addressed to nico.wunderling@pik-potsdam.de or donges@pik-potsdam.de} \and Jonathan Krönke\inst{1,2} \and Valentin Wohlfarth\inst{1,3} \and Jan Kohler\inst{1,4} \and Jobst Heitzig\inst{5} \and Arie Staal\inst{6,7} \and Sven Willner\inst{5} \and Ricarda Winkelmann\inst{1,2} \and Jonathan F. Donges\inst{1,7,}\footnotemark[1]}

\institute{Earth System Analysis, Potsdam Institute for Climate Impact Research (PIK), Member of the Leibniz Association, 14473 Potsdam, Germany \and Institute of Physics and Astronomy, University of Potsdam, 14476 Potsdam, Germany \and Department of Physics, Humboldt University of Berlin, 12489 Berlin, Germany \and Institute for Theoretical Physics, University of Leipzig, 04103 Leipzig, Germany \and Complexity Science, Potsdam Institute for Climate Impact Research (PIK), Member of the Leibniz Association, 14473 Potsdam, Germany \and Department of Environmental Sciences, Copernicus Institute of Sustainable Development, Utrecht University, Utrecht, 3584 CB, The Netherlands \and Stockholm Resilience Centre, Stockholm University, Stockholm, SE-10691, Sweden}

\abstract{Tipping elements occur in various systems such as in socio-economics, ecology and the climate system. In many cases, the individual tipping elements are not independent from each other, but they interact across scales in time and space. To model systems of interacting tipping elements, we here introduce the \emph{PyCascades} open source software package for studying interacting tipping elements (\href{http://dx.doi.org/10.5281/zenodo.4153102}{doi: 10.5281/zenodo.4153102}). \emph{PyCascades} is an object-oriented and easily extendable package written in the programming language Python. \textcolor{black}{It allows for investigating under which conditions potentially dangerous cascades can emerge between interacting dynamical systems, with a focus on tipping elements.} With \emph{PyCascades} it is possible to use different types of tipping elements such as double-fold and Hopf types and interactions between them. \emph{PyCascades} can be applied to arbitrary complex network structures and has recently been extended to stochastic dynamical systems. This paper provides an overview of the functionality of \emph{PyCascades} by introducing the basic concepts and the methodology behind it. In the end, three examples are discussed, showing three different applications of the software package. First, the moisture recycling network of the Amazon rainforest is investigated. Second, a model of interacting Earth system tipping elements is discussed. And third, the \emph{PyCascades} modelling framework is applied to a global trade network.}

\maketitle

\section{Introduction}
\label{sec:intro}
In the recent years complex systems research has increasingly focused on the matter of tipping points~\cite{gladwell2006tipping,van2016you,milkoreit2018defining} since they occur in many different systems including ecosystems, over economics, the Earth's climate system and social systems\textcolor{black}{~\cite{lenton2008tipping,may2008ecology,helbing2015saving,kopp2016tipping,tabara2018positive,otto2020social,winkelmann2020social}}. Tipping points are the critical thresholds of tipping elements, where a small perturbation can be sufficient to invoke a qualitative change of the whole system. Whether such qualitative changes can be seen as something desirable or undesirable depends a lot on the context: for instance, a potential transition of climate tipping elements towards a potential ``hothouse'' state might be dangerous for humanity~\cite{steffen2018trajectories,lenton2019climate}, while a rapid transition towards a sustainable future lies well within the scope of desired tipping events~\cite{lenton2020tipping}. However, oftentimes tipping elements do not exist in isolation, but interact across scales in time and space~\cite{rocha2018cascading,brummitt2015coupled} such as connected lakes in ecology~\cite{vangerven2017regime,scheffer2001catastrophic}, in the adoption of new technologies in the economy~\cite{comin2004cross} or the climate tipping elements in the Earth system~\citep{Kriegler2009imprecise}. Since several decades, networks are an established tool for the description of complex systems~\cite[e.g.,][]{newman2003structure,albert2002statistical}. Complex networks are structures that represent certain entities as their nodes and their interaction as their edges. They have been used, for example, to model oscillators in power grids~\cite{zou2013reviving}, food webs~\cite{gross2009generalized}, interactions of climate system components~\cite{donges2009backbone} and the collaboration network of scientists~\cite{newman2001structure}. \textcolor{black}{Critical behaviour has also been revealed on the network level. For instance, it has been shown that the likelihood of developing diabetes depends of the criticality of excitable tissue in the Langerhans Isles of the pancreas~\cite{stovzer2019heterogeneity}.}

Since there is increasing interest in modelling interacting tipping elements within the context of complex systems~\cite{gaucherel2017potential,eom2018resilience,klose2020emergence}, we bring these two strands of research together since tipping elements on networks can not only tip themselves but also imply tipping of neighbouring systems or even the network as a whole. Building upon recent developments in studying interacting nonlinear dynamics on complex networks~\cite{kronke2020dynamics,wunderling2020interacting,wunderling2020motifs,wunderling2020basin,wunderling2020network}, we here introduce the unified Python package \emph{PyCascades}. 

In Chapt.~\ref{sec:methods}, we describe \textcolor{black}{how \emph{PyCascades} can be installed and what the package contains (Sect.~\ref{subsec:install})}. Further, we describe the general structure of our package (Sect.~\ref{subsec:struc_pycascades}), the building blocks of nonlinear dynamical systems, namely the tipping elements and their interaction structure (Sect.~\ref{subsec:hopf_cusp}) as well as the network types natively included in the package (Sect.~\ref{subsec:networks}) and lastly, the extension to several types of stochastic tipping elements (Sect.~\ref{subsec:stochastic}). Thereafter, we apply our modelling framework to three different examples (Chapt.~\ref{sec:app}). First, we use our model to simulate tipping cascades in the Amazon rainforest, which is connected by a network of atmospheric moisture flows (Sect.~\ref{subsec:amazon}). Second, we show how \emph{PyCascades} can be extended to large scale Monte Carlo ensemble studies such that many uncertainties can be propagated (Sect.~\ref{subsec:climate}). Third, we exchange the fundamental differential equation that has been used in the two earlier examples to model tipping cascades in an economic example of a global trade network (Sect.~\ref{subsec:ITN}). Lastly in Chapt.~\ref{sec:conclusion}, we shortly summarise the functionalities of \emph{PyCascades}.

\section{Methods}
\label{sec:methods}
This chapter describes the basic features that are supplied by \emph{PyCascades} from the installation and the structure of the package to the fundamental features that have been developed. Here, a tutorial can be found that guides the interested reader through the most important first steps to simulate tipping cascades on interacting tipping elements (\href{http://dx.doi.org/10.5281/zenodo.4153102}{doi: 10.5281/zenodo.4153102}). Furthermore, the code for each of the following fundamental features and the three applications is provided there.

\textcolor{black}{
\subsection{Installation and package structure}
\label{subsec:install}
\emph{PyCascades} can be installed via the command line using the pip-command}

\begin{center}
    \textcolor{black}{\texttt{pip install pycascades==1.0.1}\footnote{The current version of \emph{PyCascades} is stored at https://pypi.org/project/pycascades/.}.}
\end{center}

\textcolor{black}{Alternatively, the package can directly be downloaded via the website following the zenodo-doi: 10.-5281/zenodo.4153102. The layout of the file structure of \emph{PyCascades} can be found in Tab.~\ref{tab:one}. Important files, which led to the outcomes of this work, are listed and described there. A dedicated tutorial has been developed, guiding the interested reader through some important first steps and features of the software package. For the Amazon rainforest application and the climate tipping elements application, further readme-files have been added in the respective directory. There, it is explained how the respective simulations can be started and evaluated. Additionally, the plot scripts for these two applications are deposited.}

\newpage
\thispagestyle{empty} 
\begin{landscape}
\begin{table}
\caption{\textcolor{black}{File structure of \emph{PyCascades}. Only important files are described below. Separate plot-files and extended readme-files for the Amazon rainforest and the climate tipping cascades are also supplied in the corresponding directory to facilitate the usage of \emph{PyCascades}.}}
\begin{center}
\begin{tabular}{lll}
\hline\noalign{\smallskip}
Directory & Important File(s) & Purpose \\
\noalign{\smallskip}\hline\noalign{\smallskip}
pycascades/examples & tutorial.py & Introduction to basic features\\
pycascades/examples & example\_cusp\_hopf.py & Timelines of tipping events (Fig.~\ref{fig:two})\\
pycascades/examples & network\_types\_plot.py &  Network plots (Fig.~\ref{fig:three})\\
pycascades/examples & network\_tipping\_cascade\_plot.py & Network of tipping cascades (Fig.~\ref{fig:three})\\
pycascades/examples & example\_cusp\_noise.py & Different types of noise (Fig.~\ref{fig:four})\\
pycascades/examples & economic\_cascade.py & Economic cascades\\
\noalign{\smallskip}\hline
pycascades/modules & core/gen/utils & Core of \emph{PyCascades} (Sect.~\ref{subsec:struc_pycascades})\\
\noalign{\smallskip}\hline
pycascades/amazon\_rainforest & r\_crit\_unstable\_amazon.py & Start Amazon rainforest simulations\\
pycascades/earth\_system & Main\_earth\_system.py & Start climate tipping events simulations\\
pycascades/sdeint & readme\_sdeint.txt & How to implement tipping elements with noise and \emph{sdeint}\\
\noalign{\smallskip}\hline
\end{tabular}
\end{center}
\label{tab:one}
\end{table}
\end{landscape}
\newpage

\subsection{Structure of the core of \emph{PyCascades}}
\label{subsec:struc_pycascades}
\emph{PyCascades} provides a convenient framework to solve differential equations on complex networks, i.e., it describes the dynamics of states of nodes in such a network as well as their interactions. The basic assumption is that the dynamics of tipping elements can be separated into one part for the isolated dynamics of the tipping element and another part representing the interaction terms (see Sect.~\ref{subsec:hopf_cusp} for more details). For that, it builds on \emph{SciPy} differential equation solvers~\citep{2020SciPy-NMeth} for the dynamics and on \emph{NetworkX}~\citep{hagberg2008exploring} to generate the underlying network.\\

\begin{figure}
\centering
\resizebox{\columnwidth}{!}{\includegraphics{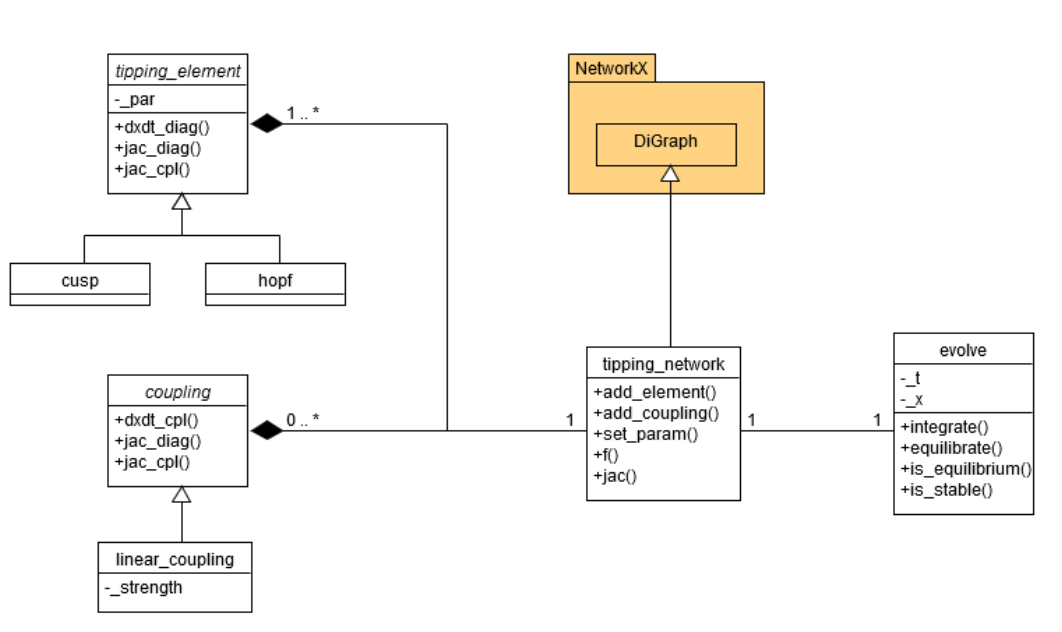}}
\caption{UML class diagram of the core of \emph{PyCascades} \textcolor{black}{that depicts structure and dependencies of \emph{PyCascades'} functionalities separated in the different python classes}. The class \texttt{tipping\_network} is derived from the \texttt{DiGraph} class of the \emph{NetworkX} package~\citep{hagberg2008exploring}. It aggregates instances of the general classes \texttt{tipping\_element} and \texttt{coupling}. The \texttt{evolve} class is associated with one instance of the \texttt{tipping\_network} class and simulates the evolution of the complex dynamical system which is implemented by the concrete \texttt{tipping\_element} and \texttt{coupling} objects with their specific parameters. For simplicity, only classes and class members important to the understanding of the \emph{PyCascades} core are shown.}
\label{fig:one}
\end{figure}

The core of \emph{PyCascades} is structured as follows (see Fig.~\ref{fig:one}). It provides the two classes \texttt{tipping\_element} and \texttt{coupling} that implement the two described types of dynamics. From these classes that can be viewed as references, concrete classes for tipping elements and interactions can be derived. Currently, \emph{PyCascades} provides the classes \texttt{cusp} and \texttt{hopf} derived from \texttt{tipping\_element} and \texttt{linear\_coupling} derived from \texttt{coupling}. Other types of tipping elements or couplings can be implemented in an analogous way. The class \texttt{tipping\_network} which is derived from the \texttt{DiGraph} class of \emph{NetworkX} is used to combine different \texttt{tipping\_element} and \texttt{coupling} objects into a network and identify each \texttt{tipping\_element} object with a node and each \texttt{coupling} object with a link. Finally, an \texttt{evolve} class is provided with methods to integrate the resulting ODE system or to trigger tipping events.

\subsection{Different types of tipping elements and interactions}
\label{subsec:hopf_cusp}
Through the \texttt{tipping\_element} class in \emph{PyCascades} different types of tipping elements can be defined and coupled together. Each tipping element can be described by its individual dynamics $f_i$ and the interaction term $g_i$, i.e., the coupling to other tipping elements. This yields
\begin{align}
\begin{split}
    \tau_i \frac{dx_i}{dt} &= f_i(x_i) + g_i(x),
\label{eq:basic_dynamics}
\end{split}
\end{align}
where $x_i$ represents the state of the respective tipping element. $\tau_i$ stands for a typical timescale of tipping. The direct interaction term $g_i(x)$ is assumed to be separable into the summands
\begin{equation}
    g_i(x) = \sum_j g_{ij}(x_i,x_j),
\label{eq:gen_cpl}
\end{equation}
linking the tipping elements $i$ and $j$.\\

\begin{figure}
\centering
\resizebox{\columnwidth}{!}{\includegraphics{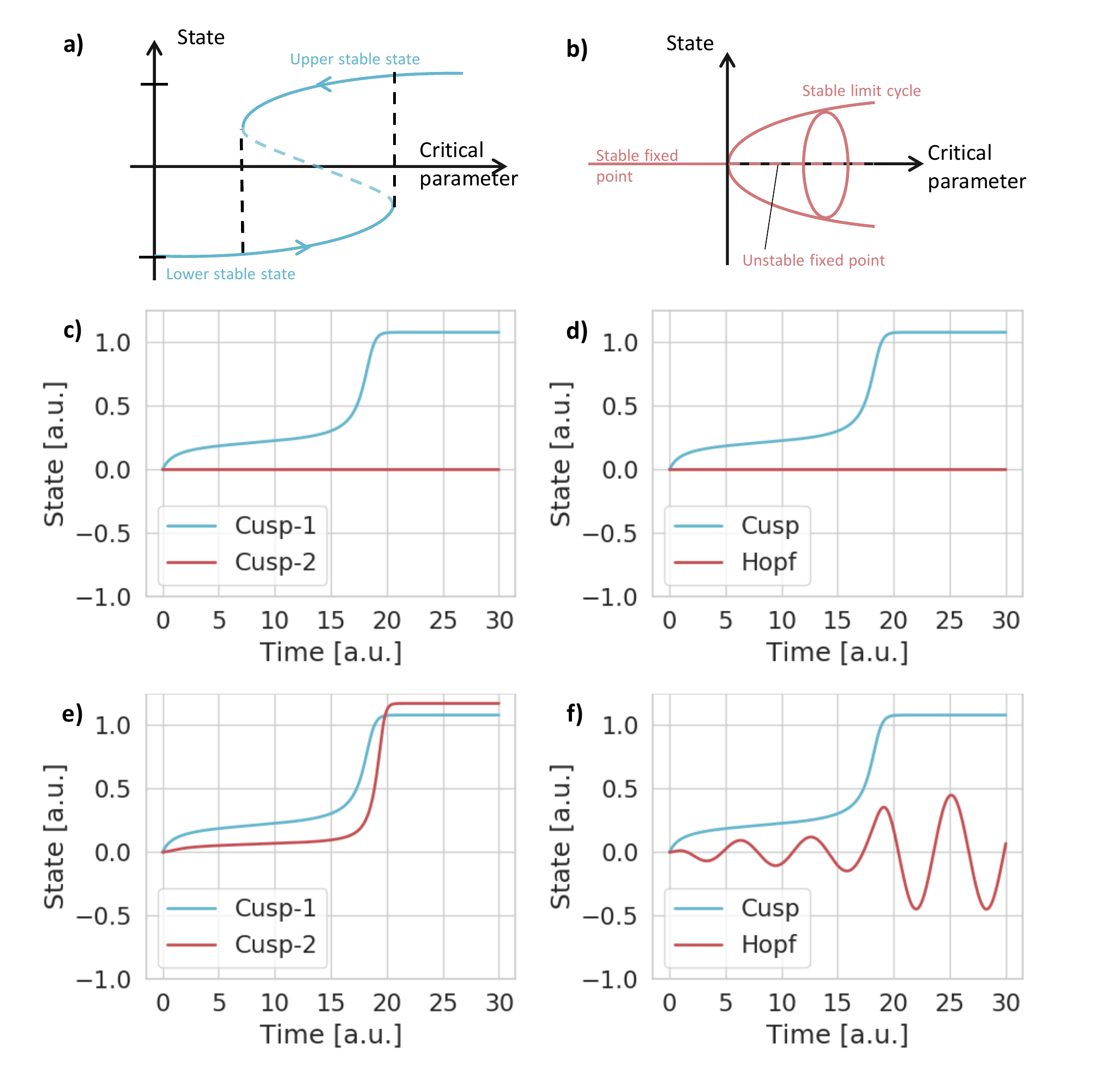}}
\caption{\textcolor{black}{The bifurcation diagrams of the pre-implemented tipping elements are shown, which possess a Cusp- or a Hopf-bifurcation, respectively (panels a) and b)). Further, two examples of tipping events are shown, once for the case where no tipping cascade emerges (panels c) and d)), and once where a tipping cascade emerges (panels e) and f) ).} \textbf{a)} Bifurcation diagram of a fold-bifurcation (see Eq.~\ref{eq:cusp}). \textbf{b)} Bifurcation diagram of a (supercritical) Hopf-bifurcation (see Eq.~\ref{eq:hopf}). \textbf{c)} Two cusp differential equations with the parameters $a_1 = a_2 = 4$, $b_1 = b_2 = 1$, $x_{0,\ 1} = x_{0,\ 2} = 0.5$ and $c_1 = 0.2$, $c_2 = 0$. Thus, the first tipping element is slightly pushed over its upper critical value and a state transition occurs. \textbf{d)} One cusp, initialised as in panel c), and one tipping element that possibly could undergo a Hopf-bifurcation ($a = 1$, $\mu = -1$). In the panels c) and d), there is no interaction between the tipping elements, i.e., $A_{ij} = 0$ $\forall i,j$. \textbf{e) and f)} Same as in panel c) and d), but with $A_{21} = 0.5$ such that the second tipping element (Cusp-2, Hopf) is coupled to the state of the first element (Cusp-1, Cusp). Therefore, in the lower panels a tipping cascade of two fold-bifurcations or, respectively, a tipping cascade of one fold- and one Hopf-bifurcation can be observed.}
\label{fig:two}
\end{figure}

In principle, any kind of tipping element can be supplied in the \texttt{tipping\_element} class of \emph{PyCascades}, but as of now, there are two kinds of tipping elements predefined \textcolor{black}{that are ready to be used and implemented. These two tipping elements are elements that possess a Cusp-bifurcation or a Hopf-bifurcation~\citep{kuznetsov2013elements}. The first pre-implemented tipping element is the Cusp-differential equation, which has been used in many contexts before to model nonlinear transitions between two alternative stable states~\cite{brummitt2015coupled,abraham1991computational}}. The normal form of its differential can be written as
\begin{equation}
    f_\text{Cusp} (x) = \frac{dx}{dt} = -a\left(x-x_0\right)^3 + b\left(x-x_0\right) + c.
\label{eq:cusp}
\end{equation}
Here, $a, b > 0$ and $x_0$ represents a shift on the x-axis. The parameter $c$ is the critical parameter, which invokes a shift from a lower stable state to an upper stable state as soon as the critical value $c_\text{crit,\ high}$ is surpassed. The other way round, when $c$ is diminished, a state transition from the upper to the lower stable state occurs at $c_\text{crit,\ low}$. Eq.~\ref{eq:cusp} has the normal form of a fold-bifurcation and has, as a paradigmatic model, been applied in many different areas such as systems in ecology, climate science and economics~\cite{brummitt2015coupled,klose2020emergence,wunderling2020interacting,van2007theory,scheffer2007regime}. For the special case that $a=4, b=1$ and $x_0=0.5$, the two stable states are located at $x_{1}= 0$ and $x_{2} = 1$ for $c=0$. The critical parameter lies at $c_\text{crit,\ high} = - c_\text{crit,\ low} = \sqrt{(4b^3)/(27a)} = \sqrt{4/(27\cdot4)} \approx 0.19$. The bifurcation diagram of this equation is shown in Fig.~\ref{fig:two}a.

The second tipping element that is provided by \emph{PyCascades} is a Hopf-bifurcation. The normal form in polar coordinates of this bifurcation can be written as 
\begin{equation}
\begin{split}
    f_\text{Hopf,\ r}(r) = \frac{dr}{dt} &= \left(\mu - r^2\right)\cdot r a \\
    f_{\text{Hopf,\ } \phi} (\phi ) = \frac{d \phi}{dt} &= b
\label{eq:hopf}
\end{split}
\end{equation}
with the parameters $a$ and $b$. Here, the Hopf-bifurcation is given in polar coordinates with the radius $r$ and the angle $\phi$. Importantly, $\mu$ is the critical parameter and a bifurcation from a stable fixed point to a stable limit and an unstable fixed point occurs when $\mu$ crosses zero from below. The bifurcation diagram is shown in Fig.~\ref{fig:two}\textcolor{black}{b}. Applications of Hopf-bifurcations have been found, for instance, in predator-prey cycles in Lotka-Volterra systems or in the Hodgkin-Huxley model of neurons~\cite{gardini1989hopf,guckenheimer1993bifurcation}. In the climate system, there exist conceptual models that represent the El-Ni\~{n}o Southern Oscillation as a Hopf bifurcation~\cite{timmermann2003nonlinear,dekker2018cascading} based on a model by Zebiak \& Cane (1987)~\cite{zebiak1987model}.\\

Next, for the interactions, any type of coupling can in principle be used and implemented in \emph{PyCascades}. However, for the moment only linear interactions are considered
\begin{equation}
    g_i(x) = \sum_{j=1}^N A_{ij} x_j.
\label{eq:interaction}
\end{equation}
If there is a connection between tipping element $i$ and $j$, then $A_{ij} \neq 0$, otherwise $A_{ij} = 0$. In Fig.~\ref{fig:two}c-f, we show an example how tipping cascades can emerge from the coupling between two tipping elements for the case of two cusp-differential and for the case of one cusp coupled to the normal form of a Hopf-bifurcation.

\subsection{Paradigmatic network types of interacting tipping elements}
\label{subsec:networks}
\begin{figure}
\centering
\resizebox{\columnwidth}{!}{\includegraphics{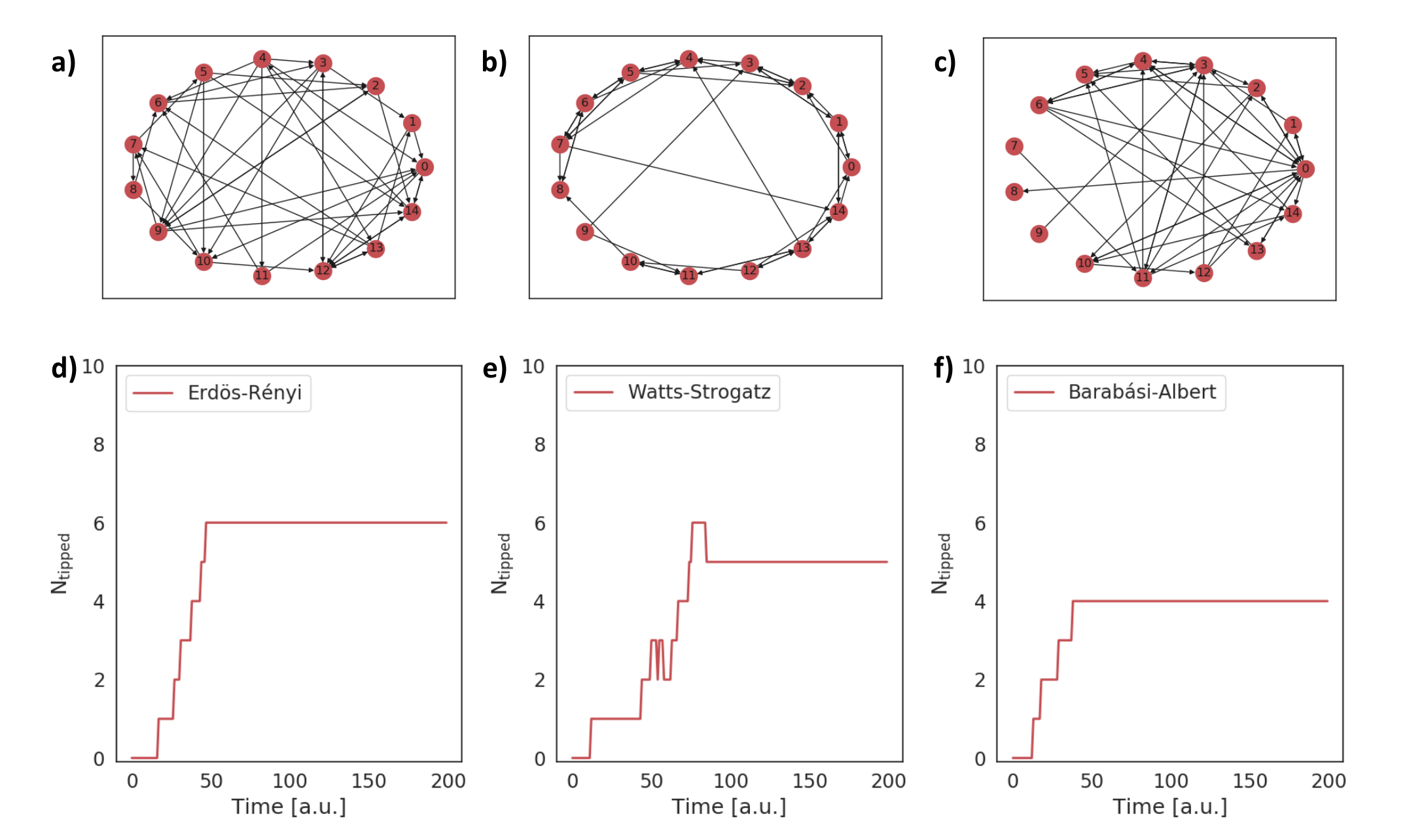}}
\caption{\textcolor{black}{The three paradigmatic and pre-implemented network types are shown together with an example of a tipping cascade in this network.} Exemplary network structure with 15 nodes and an average degree of 3 for \textbf{a)} an Erd\H{o}s-Rényi network, \textbf{b)} a Watts-Strogatz network (rewiring probability: 0.15) and \textbf{c)} a Barabási-Albert network. Exemplary tipping cascades for a network of 15 nodes, an average degree of 3, where each node is represented by a Cusp-tipping element (see Eq.~\ref{eq:cusp}) with $a=4$, $b=-1$, $x_0=0.5$ for all nodes. The couplings between the nodes are alternately set to 0.2 and $-$0.4 for interaction 1, 2, 3, etc.. Then, at $t=0$ one randomly chosen node $i$ is set to the upper state by choosing $c_i=0.2$ such that tipping cascades can emerge. The number of tipped elements are shown for \textbf{d)} an Erd\H{o}s-Rényi network, \textbf{e)} a Watts-Strogatz network (rewiring probability: 0.15) and \textbf{f)} a Barabási-Albert network.}
\label{fig:three}
\end{figure}

With \emph{PyCascades} it is possible to investigate the dynamics of tipping and tipping cascades in larger directed networks. These types of networks can either be explicitly spatially embedded (see Chapt.~\ref{sec:app}) or well-known predefined network models such as the Erd\H{o}s-Rényi model, the Barabási-Albert model or the Watts-Strogatz model~\cite{erdos1959random,barabasi1999emergence,watts1998collective}. Originally, the network models that are inbuilt in python's network package \emph{NetworkX} are undirected for Watts-Strogatz networks and Barabási-Albert networks, while we require directed networks. Additionally, it might also be helpful or necessary to be able to determine a certain average degree. Therefore, a generalisation of these two networks types has been developed. (i) Watts-Strogatz network: A regular network is created where each node $i$ is connected to its $m$ closest neighbours in both directions. $m$ must be an even integer and the average degree $\left< k \right> = m$. $m$ is chosen in such a way that the average degree of the resulting network is larger than the desired average degree. Then, links are randomly deleted until the desired average degree is reached. Lastly, each of the remaining links is rewired with the desired rewiring probability as in the usual Watts-Strogatz model. (ii) Barabási-Albert model: First, two nodes are bidirectionally coupled. Each further node is, again, bidirectionally coupled to one already existing node $i$ with the probability $p = (k^{\text{in}}_{i} + k^{\text{out}}_i)/(\sum_{mn} a_{mn})$, where $k^{\text{in}}_{i}$ is the in-degree and $k^{\text{out}}_i$ is the out-degree of node $i$. With $a_{mn}$, the sum of all edges in the network is denoted. In the end, the average degree $\left< k \right>$ depends on the stochastic network. Therefore, it can happen that the actual average degree $\left< k \right>$ of the network exceeds or \textcolor{black}{falls below} the desired average degree $k_\text{des}$. While $\left< k \right> < k_\text{des}$, another link is added between two randomly selected nodes $i$ and $j$. While $\left< k \right> > k_\text{des}$, a link between two randomly selected nodes $i$ and $j$ is deleted. For comparison of the construction of these network models, see also Krönke et al.(2020)~\cite{kronke2020dynamics}. Examples for a realisation of these three network types and an exemplary tipping cascade in those can be found in Fig.~\ref{fig:three}.

\subsection{Stochasticity in tipping elements}
\label{subsec:stochastic}
In the real-world, systems often underlie fluctuations, which under certain circumstances can cause critical transitions, called \textit{noise-induced tipping}. Numerous prominent examples can be found in dynamical systems such as in electronics, optics or neurons, but also in ecology and in the Earth's climate system~\cite{kondepudi1986observation,gammaitoni1998stochastic,scheffer2009early,thompson2011predicting,ashwin2012tipping}. Therefore, we decided to create a class for a stochastic version of the cusp tipping element (Eq.~\ref{eq:cusp}) for additive noise
\begin{equation}
   \textcolor{black}{f_\text{Cusp,\ stoch.} = dx = \left[ -a\left(x-x_0\right)^3 + b\left(x-x_0\right) + c \right] dt + \sigma dW.}
\label{eq:cusp_stochastic}
\end{equation}
Here, $\sigma$ denotes the noise level and $dW/dt$ describes the Wiener process or Brownian motion. In the case of random white noise (Gaussian white noise) as used here, $W$ is sampled from a Gaussian distribution. To implement stochastic differential equations, python's SciPy function \texttt{odeint} has been replaced by \texttt{sdeint}~\cite{aburn2017sdeint}. \texttt{sdeint} has several algorithms implemented, which are able to solve stochastic differential equations. Here, and in the provided version of \emph{PyCascades}, an order 1.0 strong stochastic Runge-Kutta algorithm is employed~\cite{rossler2010runge}.

\begin{figure}
\centering
\resizebox{\columnwidth}{!}{\includegraphics{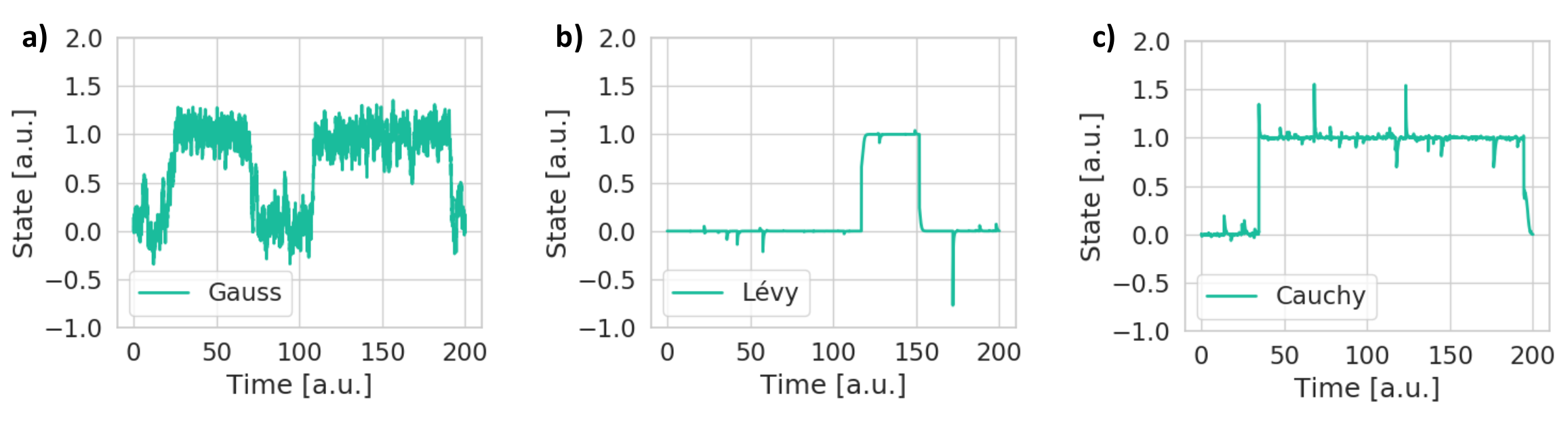}}
\caption{\textcolor{black}{Noise-induced tipping events with respect to different types of noise (Gauss, Lévy, Cauchy) that are available in \emph{PyCascades}.} Simulation of one Cusp-tipping ($a=4$, $b=1$, $c=x_0=0$) element with stochastic noise (see Eq.~\ref{eq:cusp_stochastic}) of the following type: \textbf{a)} Gaussian noise ($\sigma=0.25$, see Eq.~\ref{eq:stat_distributions}), \textbf{b)} Lévy noise ($\sigma = 1.0$, see Eq.~\ref{eq:stat_distributions}) and \textbf{c)} Cauchy noise ($\sigma = 1.0$, see Eq.~\ref{eq:stat_distributions}).}
\label{fig:four}
\end{figure}

Furthermore, Gaussian noise distributions are not necessarily able to describe all types of fluctuations in real-world systems since in reality noise might be correlated or not be standard normally distributed. Besides Gaussian noise, \emph{PyCascades} allows to compute systems with Lévy and Cauchy noise (see Fig.~\ref{fig:four}). These types of noise (Lévy, Cauchy) may be more suitable for describing extreme events than Gaussian noise, however,  in the implemented form they still remain uncorrelated. It has been found that the probability of jumping between the two stable states in a double-well potential is impacted by single strong extreme events from those $\alpha$-stable noise distributions~\cite{ditlevsen1999anomalous}. For instance, it has been proposed that this might have been of relevance for climate system states on a millennial time scale during the last glacial period as was observed in ice-cores~\cite{ditlevsen1999observation}. Also, transitions triggered by extreme events emerging from Lévy-distributions in other nonlinear climate system components such as the Amazon rainforest or the thermohaline circulation have been investigated, as well as transitions in gene expression processes in molecular biology~\cite{tesfay2020influence,serdukova2017metastability,zheng2016transitions}. 

The distributions for Gaussian, Lévy and Cauchy noise in \emph{PyCascades} are taken from python's SciPy libraries
\begin{equation}
\begin{split}
    p_\text{Gauss}(x) &= \frac{1}{\sqrt{2 \pi \sigma^2}} \cdot \exp \left( - \frac{x^2}{2 \sigma^2} \right) \\
    p_\text{Lévy}(x) &= 0.5\cdot p_\text{Lévy,\ pos.} + 0.5 \cdot p_\text{Lévy,\ neg.} = \\ &= 0.5 \cdot \sqrt{\frac{\sigma}{2 \pi x^3}} \cdot \exp \left( -\frac{\sigma}{2x} \right) + 0.5 \cdot \sqrt{\frac{\sigma}{2 \pi \left| x \right|^3}} \cdot \exp \left( -\frac{\sigma}{2\left| x \right|} \right) \\
    p_\text{Cauchy}(x) &= \frac{1}{\sigma \pi} \cdot \frac{\sigma^2}{\sigma^2 + x^2},
\end{split}
\label{eq:stat_distributions}
\end{equation}
where $\sigma$ is the standard deviation and the mean value $\mu \equiv 0$.

\section{Applications}
\label{sec:app}
In this section, we show three examples of how \emph{PyCascades} can be applied to real-world systems. The first application is the moisture-recycling network of the Amazon rainforest, where we introduce \emph{PyCascades} on a spatially embedded network. In the second application in a subset of interacting climate tipping elements, we combine the \emph{PyCascades} modelling framework with a large-scale setup of Monte Carlo simulation to show how numerous uncertainties in parameters can be propagated systematically. The third application, a global trade network of more than 5~000 nodes and 400~000 links, complements our analysis by simulating tipping cascades with a modernised, economically motivated differential equation (see Eq.~\ref{eq:diffeq_ecolog_element}) replacing the Cusp-differential equation (see Eq.~\ref{eq:cusp}).

\subsection{The Amazon rainforest}
\label{subsec:amazon}
\begin{figure}
\centering
\resizebox{\columnwidth}{!}{\includegraphics{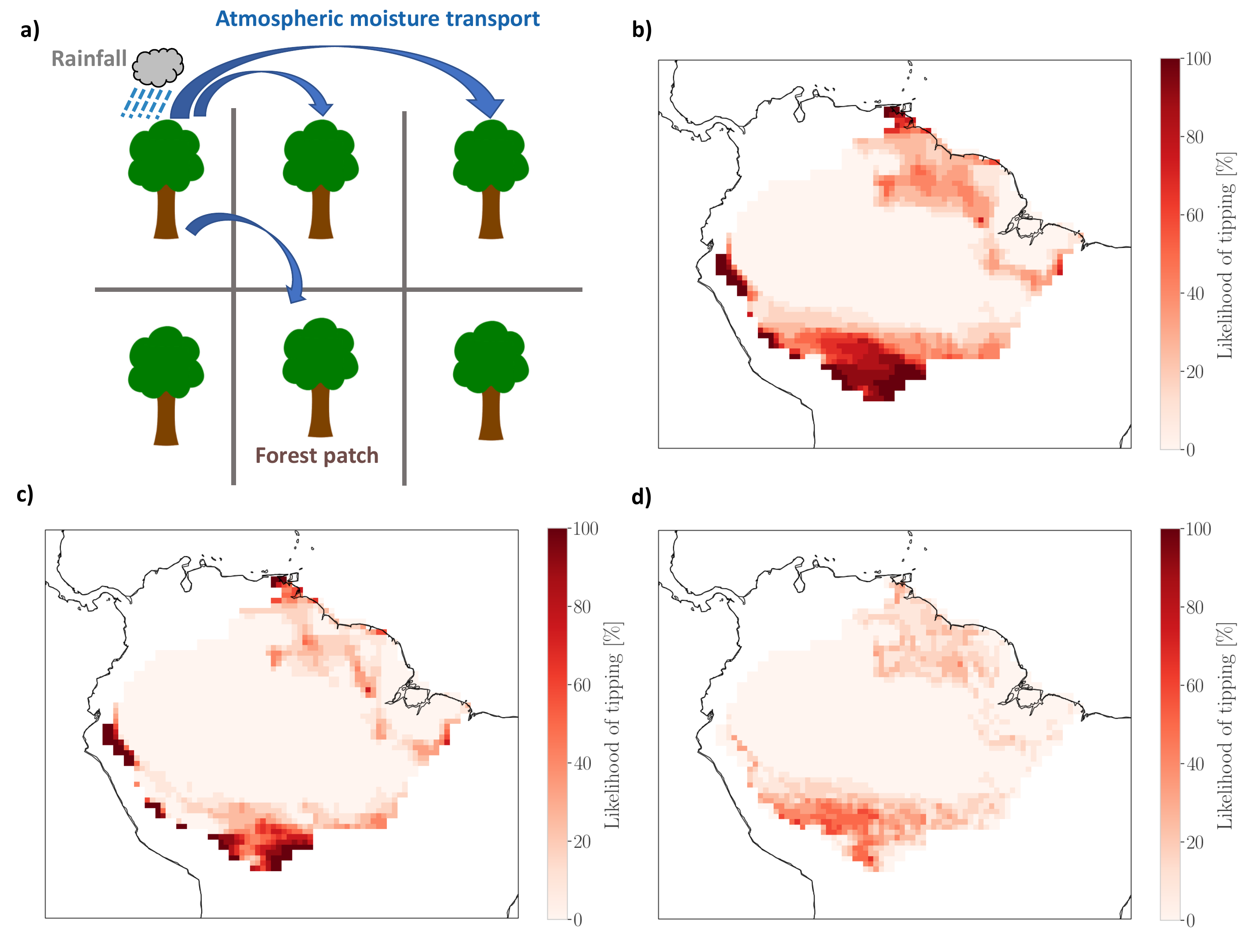}}
\caption{\textcolor{black}{Tipping cascades in a conceptual model of the Amazon rainforest connected via an atmospheric moisture recycling network}. \textbf{a)} Sketch of the network of interacting rainforest patches in the Amazon rainforest. Precipitation rains down over some parts of the rainforest and parts of it are re-evapotranspirated and transported further by the wind (atmospheric moisture transport). \textbf{b)--d)} Exemplary tipping experiment on a 0.5$\times$0.5$^\circ$ grid, where each grid cell represents one rainforest patch. The colourbar represents the likelihood of tipping averaged over the years 2003--2014. We show a comparison between \textbf{b)} coupling switched on (see Eq.~\ref{eq:amazon}), \textbf{c)} coupling switched off (see Eq.~\ref{eq:amazon} with $g_i(x) \equiv 0$ $\forall i$) and \textbf{d)} the difference between the panels b) and c). For each year in the study period (2003-2014), we performed one such tipping experiment, and the results shown are an average over this period.}
\label{fig:five}
\end{figure}

It is suspected that the Amazon rainforest is a tipping element in the Earth's climate system~\cite{lenton2008tipping}, which might approach a tipping point due to various anthropogenic pressures including climate change, fires and land-use change~\cite{nobre2016land,davidson2012amazon,cox2008increasing}. The Amazon rainforest might exhibit multistability at certain rainfall levels, as suggested by conceptual models and observational data~\cite{zemp2017self,staal2015synergistic,van2014tipping,hirota2011global,staver2011global}. This implies that rainforest patches may transition to a savannah-like state when the rainfall drops below a certain critical level. These rainforest patches depend on each other, as rain is re-evaporated by the trees and thus preserved in the system through atmospheric moisture recycling~\cite{aragao2012environmental,eltahir1994precipitation} (see Fig.~\ref{fig:five}a). This means that the Amazon rainforest is an excellent example of how tipping cascades can travel through a system, which can be modelled with \emph{PyCascades}. We divide the Amazon into 0.5$\times$0.5$^\circ$ (appr. 50~km) grid cells and assume that each is an interacting tipping element that can be described by the Eqs.~\ref{eq:cusp} and~\ref{eq:interaction}. For simplicity, we chose $a_i=b_i=1$ and $x_{0,i}=0$ for all tipping elements and further assume that the critical parameter is only dependent on the rainfall a rainforest cell receives, which tips in case the received rainfall is less than the critical rainfall. Then, the critical parameter $c_i$ and the coupling $g_i(x)$ can be denoted as
\begin{equation}
\begin{split}
    c_i &= c_0 \cdot \frac{R_i - \left< R \right>}{R_\text{crit} - \left< R \right>} \\
    g_i(x) &= \frac{1}{2} \frac{c_0}{R_\text{crit} - \left< R \right>}  \sum^N_{j=1} \delta^\text{Rain}_{ij}.
\label{eq:amazon}
\end{split}
\end{equation}
Here, $R_i$ is the rainfall in cell $i$, $\left< R \right>$ is the average rainfall over the whole Amazon basin and $R_\text{crit}$ is the critical rainfall. Further, $c_0 = \sqrt{4/27}$ if $a=b=1$ and $x_0=0$. Lastly, $\delta^\text{Rain}_{ij}$ is the moisture transport in mm/yr from cell $j$ to cell $i$. Since the distance between the two stable states is 2, a prefactor of 1/2 is required to re-normalise the coupling. We choose the critical rainfall $R_\text{crit}$ to be 1700~mm/yr for all cells, which is approximately the value below which the alternative savannah state becomes more resilient than the rainforest state~\cite{staal2016bistability}. The atmospheric moisture recycling simulations used in this work were performed by Staal et al.~(2018)~\cite{staal2018forest} for the years 2003--2014 and assembled into a network by Krönke et al. (2020)~\cite{kronke2020dynamics}. In this simplified example, we assume that if a forested grid cell tips, moisture recycling via that cell stops. We performed a tipping experiment for each year between 2003 and 2014 and averaged the results over this period. We find tipping events in several parts of the Amazon basin which cascade to other forest patches (Fig.~\ref{fig:five}b--d). This analysis illustrates how \emph{PyCascades} can be applied to simulate tipping events and cascades in a real-world network of interacting tipping elements.

\subsection{Climate tipping elements}
\label{subsec:climate}
Apart from the Amazon rainforest, there exists a range of processes and systems in the Earth's climate system that exhibit threshold behaviour~\cite{lenton2008tipping}. These tipping elements contain biosphere components (e.g. Amazon rainforest, coral reefs), large-circulation patterns (e.g. Atlantic Meridional Overturning Circulation, monsoon systems) or cryosphere components (e.g. Greenland Ice Sheet, West Antarctic Ice Sheet). Under ongoing global warming, many of them are at risk of transitioning into an alternative, tipped state at lower levels of global warming than previously though~\cite{lenton2019climate,schellnhuber2016right}. Such transitions would have dangerous consequences for humanity and biosphere integrity in the Earth system~\cite{steffen2018trajectories,lenton2019climate}. There is an additional risk that tipping elements are strengthened by reinforcing, positive feedbacks within the climate system such that cascades might be triggered, potentially up to a planetary-scale tipping cascade that could push the Earth towards a ``hothouse'' state~\cite{steffen2018trajectories}. Moreover, the tipping elements in the climate system are interacting and there is a subset of five tipping elements where the interaction structure has been made explicit by an expert elicitation~\cite{Kriegler2009imprecise}. This network and their interactions have been used by some studies to investigate the risk of tipping cascades in the climate system, but also to quantify economic damages exerted by interacting tipping elements~\cite{gaucherel2017potential,wunderling2020basin,cai2016risk}.

\begin{figure}
\centering
\resizebox{\columnwidth}{!}{\includegraphics{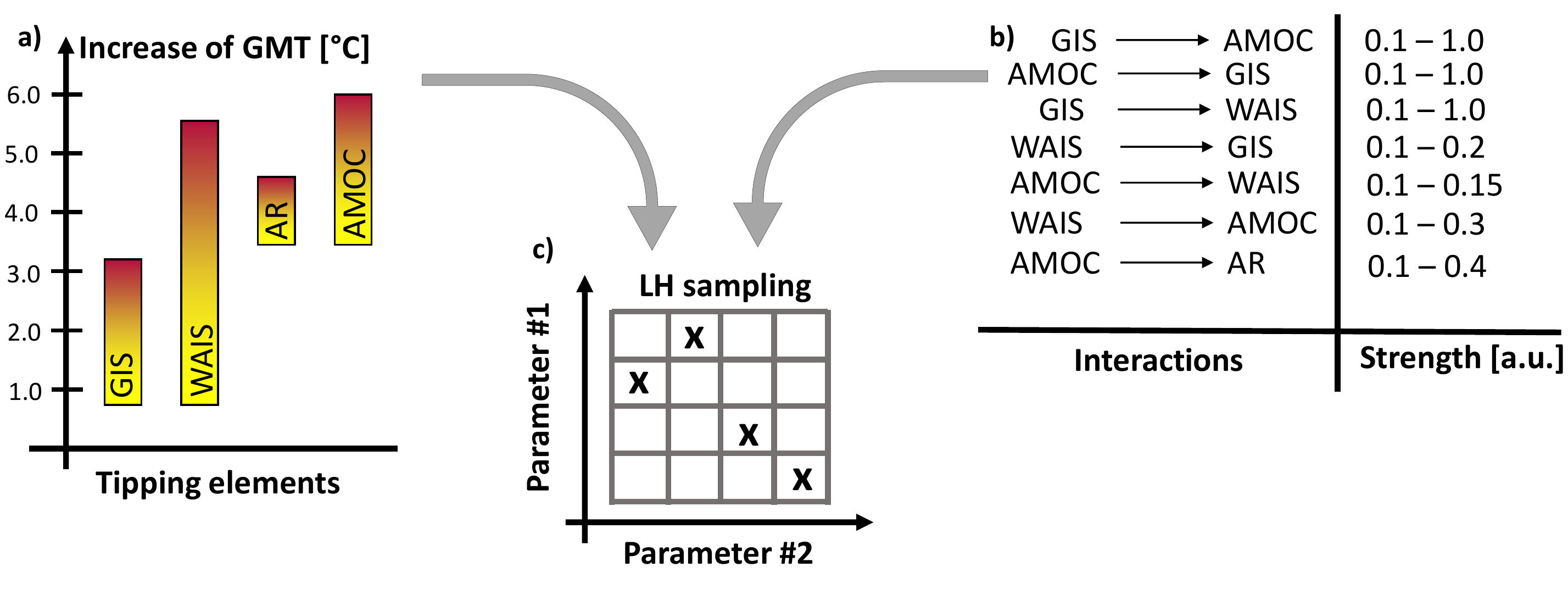}}
\caption{\textcolor{black}{Construction of the simulation ensemble to cover a large part of the phase space.} \textbf{a)} Uncertainties in the critical temperatures~\cite{schellnhuber2016right} and \textbf{b)} interaction strengths~\cite{Kriegler2009imprecise} are put into a latin hypercube sampling algorithm~\cite{pyDOE2013latin} to construct suitable initial conditions (\textbf{c)}) that cover a larger part of the state space than normal random sampling would.}
\label{fig:six}
\end{figure}

\begin{figure}
\centering
\resizebox{0.75\columnwidth}{!}{\includegraphics{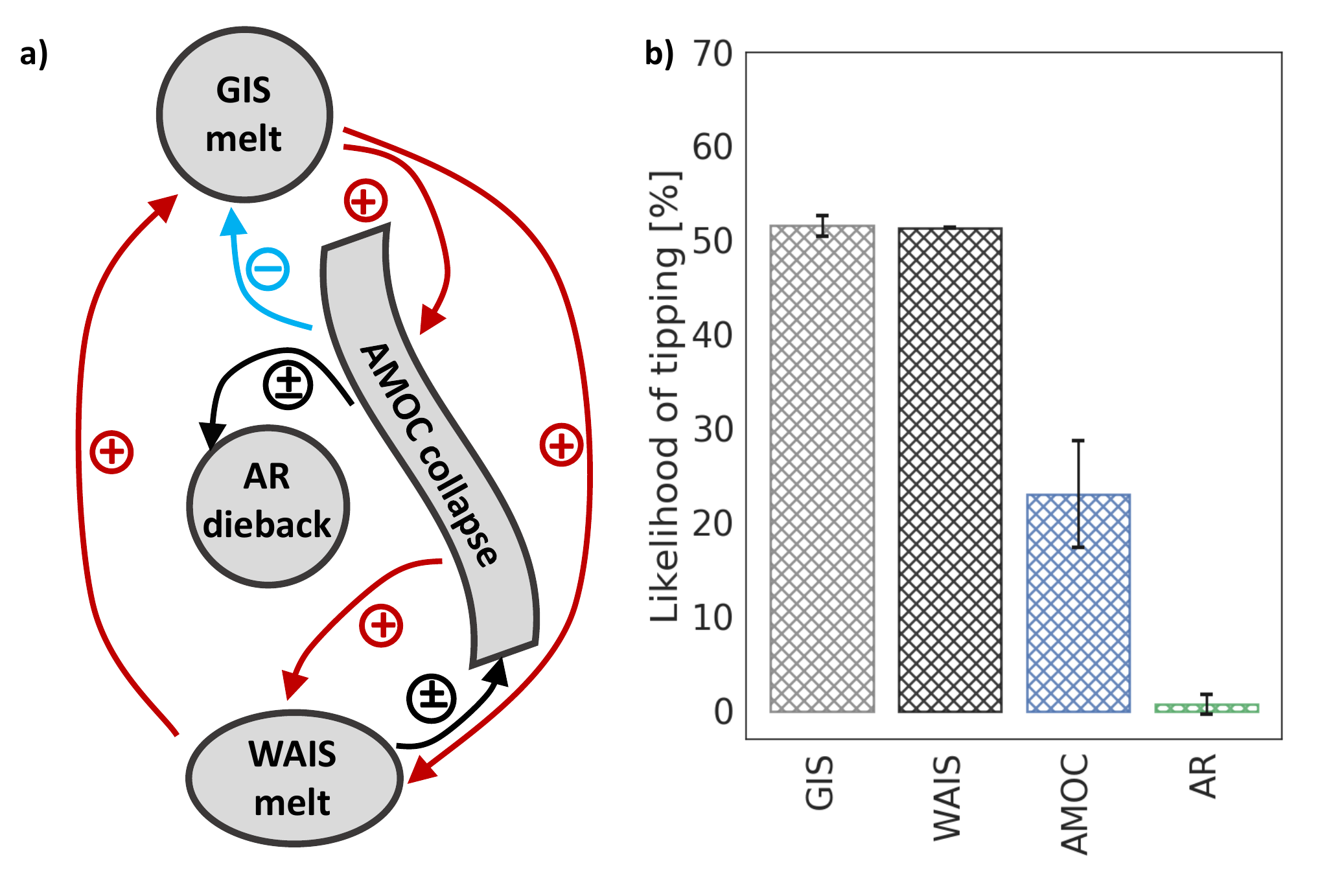}}
\caption{\textcolor{black}{Tipping cascades in a conceptual model of climate tipping elements}. \textbf{a)} Interaction structure of four tipping elements (Greenland Ice Sheet: GIS, West Antarctic Ice Sheet: WAIS, Atlantic Meridional Overturning Circulation: AMOC, AR: Amazon rainforest). Destabilising interactions are depicted in red, stabilising interactions are depicted in light blue and uncertain interactions are shown in black. \textbf{b)} Likelihood for the respective tipping element to transgress its stable branch computed by running Eq.~\ref{eq:climate_tipping_cusp} into equilibrium. The error bar shows the standard deviation arising from the nine different possibilities of constructing the network. There are two uncertain links since their direction of interaction is unclear, meaning they could be stabilising, destabilising or zero, i.e., $[-,0,+]$. Permuting these three options, gives nine different network structures and for each of them, we simulate 1000 ensemble members. We chose a scenario, where $\Delta \text{GMT} = 2$~$^\circ$C above pre-industrial levels.}
\label{fig:seven}
\end{figure}

Here, we show how \emph{PyCascades} can be used to simulate tipping events in four of the five aforementioned tipping elements: the Greenland Ice Sheet (GIS), the West Antarctic Ice Sheet (WAIS), the Atlantic Meridional Overturning Circulation (AMOC) and the Amazon Rainforest (AR). For these four tipping elements, there exist conceptual models of their nonlinear behaviour with respect to a forcing parameter~\cite{zemp2017self,van2014tipping,levermann2016simple,wood2019observable,stommel1961thermohaline}, which can be traced back to increases in levels of global warming above pre-industrial~\cite{schellnhuber2016right}. Therefore, we can arguably model these four elements by 
\begin{equation}
\begin{split}
    &\frac{dx_i}{dt} = \left[ \underbrace{- x_i^3 + x_i + \frac{\sqrt{4/27}}{T_\text{crit,\ i}} \cdot \Delta \text{GMT}}_{\text{Individual\ dynamics\ term}} + \underbrace{\sum_{\substack{j\\j \neq i}} \frac{s_{ij}}{4} \left( x_j + 1 \right)}_{\text{Interaction\ term}} \right] \frac{1}{\tau_i} \\
    &\text{with\ } i=\left\{ \text{GIS,\ WAIS,\ AMOC,\ AR} \right\}.
\end{split}
\label{eq:climate_tipping_cusp}
\end{equation}
Here, $\Delta$GMT is the increase of the global mean temperature, $T_\text{crit,\ i}$ the critical temperature at which a certain tipping element transgresses its baseline state, $s_{ij}$ the interaction strength between the tipping elements and $\tau_i$ the time a certain tipping event needs. Each $s_{ij}$ has a certain physical meaning, for instance, the freshwater entry from the GIS weakens the AMOC, while a weaker AMOC cools the northern hemisphere at the same time~\cite[e.g.][]{caesar2018observed}. The typical tipping time scales $\tau_i$ are chosen to be 4900, 2400, 300 and 50 years at 4~$^\circ$C above pre-industrial levels of global warming for GIS, WAIS, AMOC and AR, respectively. For more details, see Fig.~\ref{fig:six} and please be referred to Wunderling et al.~(2020)~\cite{wunderling2020interacting}. The parameter uncertainties and a potential interaction structure are shown in Figs.~\ref{fig:six} and~\ref{fig:seven}. In Eq.~\ref{eq:climate_tipping_cusp}, there are many parameters with uncertainties, for instance at which temperature $T_\text{crit,\ i}$ a critical transition occurs or how strong the interactions $s_{ij}$ are. While upper and lower limits are given in the literature~\cite{Kriegler2009imprecise,schellnhuber2016right}, their uncertainty need to be propagated thoroughly. For this purpose, we use a large scale Monte-Carlo ensemble based on the latin hypercube sampling (LHS) method \emph{pyDOE}~\cite{pyDOE2013latin}. The LHS is a sampling method that generates initial conditions that can be used in a Monte Carlo ensemble. They cover the state space of all uncertain parameters to a higher degree than random sample generation and are therefore better suited to create Monte Carlo ensembles in higher-dimensional systems. In Figs.~\ref{fig:six} and~\ref{fig:seven}, we demonstrate that constructing a large-scale Monte Carlo ensemble can be combined with simulating tipping cascades with \emph{PyCascades}. In the critical temperatures $T_\text{crit,\ i}$ and the interaction strengths $s_{ij}$ are 11 parameters with uncertainties (see Fig.~\ref{fig:six}). Upon that, we construct an ensemble of 1000 initial conditions. 

In Eq.~\ref{eq:climate_tipping_cusp}, we assume that the interaction term is 25\% as important as the individual dynamics term. Thus, the interaction strength $s_{ij}$ is divided by 4 in Eq.~\ref{eq:climate_tipping_cusp}. While this poses a hypothetical scenario, it allows us to estimate the likelihood of tipping of certain element at a certain increase of the global mean temperature $\Delta \text{GMT}$. For 2~$^\circ$C, we find that the likelihood of tipping is around 50\% for the GIS and WAIS, while it is significantly lower for the AMOC (around 25\%) and the AR (less than 5\%). There is a relatively high likelihood that GIS and WAIS tip since their critical temperature is lowest and there is a strong  interaction link from GIS to WAIS. Therefore, the likelihood of tipping is lower for the AMOC, but the uncertainty is higher due to the strong negative feedback loop with GIS. Lastly, the AR has a very low likelihood of tipping since it is only connected to the other tipping elements via one uncertain link from AMOC.

\subsection{International Trade Network}
\label{subsec:ITN}
In the third example, we apply the \emph{PyCascades} framework of interacting tipping elements to the \textit{International Trade Network}. We construct the network from the EORA multi-regional input-output (MRIO) database~\cite{lenzen2012ITN} as also done in other studies \cite{bierkandt2014acclimate,wenz2014acclimate}. The database, which has also been subject to static analyses~\cite{maluck2015network}, consists of 188 countries with 27 economic sectors each, and includes the annual monetary flows between these sectors and regions. We interpret the individual sectors in each country as nodes of a network, and the flow $f_{ij}$ in the MRIO table as the weight for each directed link from node $j$ to $i$. In our analysis, we use the data for the year 2012. Following previous analyses~\cite{bierkandt2014acclimate,wenz2014acclimate,otto2017modeling}, we also use a threshold of 10$^6$ US-\$ such that we exclude unrealistically small flows.

Propagation of economic losses on the trade network has previously been studied, for instance, with the \textit{Acclimate} model~\cite{otto2017modeling}. This model interprets the economic sectors in each country as \textit{firms} producing a \textit{commodity} specific for the respective sector. Each firm does so using other commodities as inputs with specific, fixed proportions according to a \textit{Leontief} production function \cite{fandel1991production} as also used in simpler input-output models. These fixed proportions are taken from the multi-regional input-output (MRIO) table underlying the construction of the trade network, which constitute the \textit{baseline state} (untipped state) of the model. If, for instance, the transportation sector in a country receives an input of ten billion US-\$ from the oil sector and 90 billion US-\$ from the machinery sector, it might have an output of 110 billion US-\$ according to the MRIO table such that the created surplus would be 10 billion US-\$. However, the according sector always produces according to these proportions using nine times as much ``machinery commodity'' as ``oil commodity'', and produces ten percent more ``transportation commodity'' than the sum of its inputs. If a firm receives only a certain fraction of the baseline state of a commodity due to some perturbation, the firm's output is limited to the same fraction of the baseline output. However, in the Acclimate model firms have idle capacities, i.e. the ability to temporarily produce more than their baseline output, if they have the necessary inputs and demand is high. The dynamics of this \textit{anomaly model} are focused on perturbations around the baseline state with each agent aiming for maximum profit and minimum costs under local circumstances. After a shock or perturbation ceases the model returns back to the baseline state, which constitutes an equilibrium of the model's dynamics.

\begin{figure}
\centering
\resizebox{\columnwidth}{!}{\includegraphics{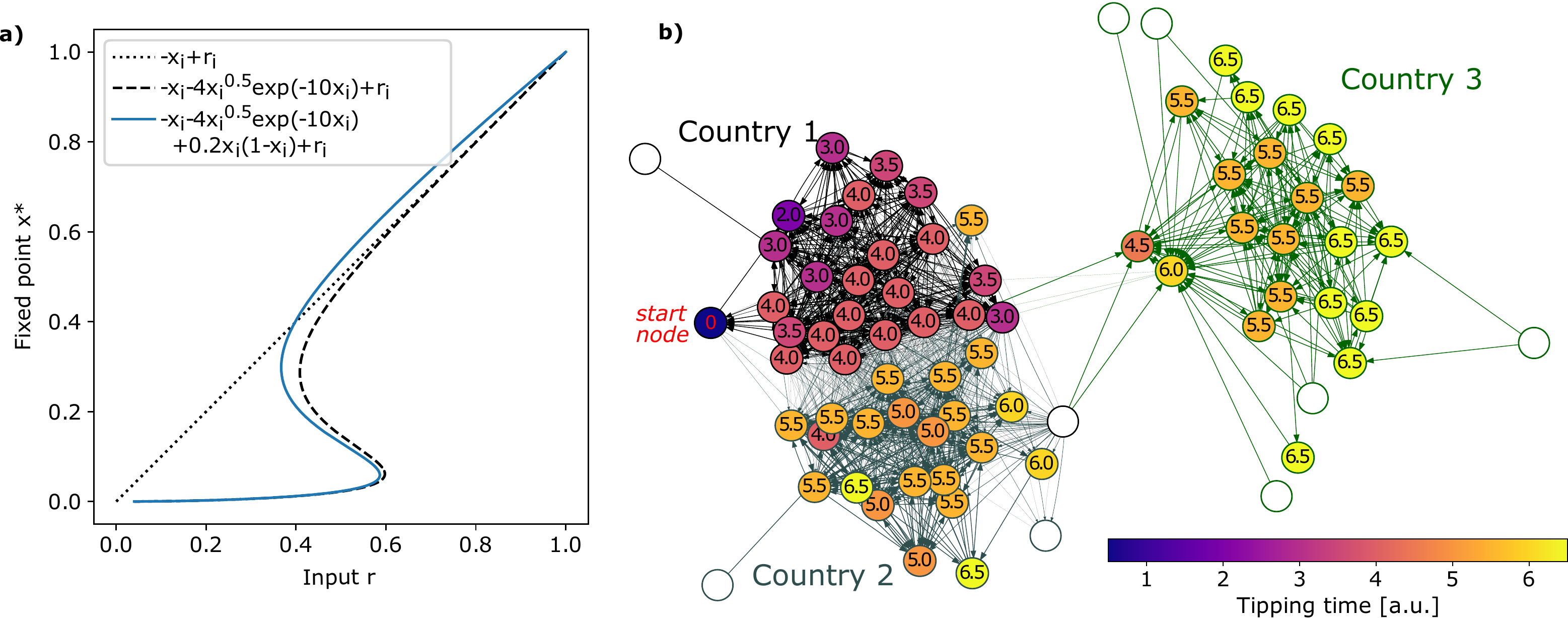}}
\caption{\textcolor{black}{Tipping cascades in a conceptual model of interacting countries in the \emph{International Trade Network}}. \textbf{a)} Bifurcation diagram for the proposed economic tipping element defined by equation \ref{eq:diffeq_ecolog_element}. \textbf{b)} Tipping cascade between three different countries starting at the red \textit{start node} at $t=0$. The colour of the nodes represents the time at which a certain sector in that country tips and the colour of the arrows indicate the targeted country.}
\label{fig:eight}
\end{figure}

Tipping is not at the centre of the Acclimate model whose scope, as an anomaly model, vanishes for very large perturbations such as bankruptcies. We here, thus, define a simple dynamic for tipping on the trade network while keeping the linear Leontief production assumption for small perturbations, whereas nonlinear dynamics are assumed for larger perturbations. Nonlinear behaviour and tipping is common in economic networks, for instance in the banking sector~\cite{may2008banking,haldane2011banking}. The nodes in the trade network are only to be perceived as representative firms, i.e., aggregates of national sectors, which consist of a variety of connected actors -- so each node represents a network itself and might show nonlinear as well as tipping behaviour. The standard form of a tipping element defined by the Cusp-differential equation (see Eq.~\ref{eq:cusp}) is not well suited for this purpose. Instead, we are here looking for a new differential equation with the following properties:
\begin{enumerate}
	\item[1)] The state $x_i$ of a node $i$ should represent its productivity that is between 0 (no production) to 1 (full production).
	\item[2)] The element should react almost linearly to small perturbations as in the Acclimate model.
	\item[3)] For large perturbations there should be a collapse of the productivity, including tipping and hysteresis.
\end{enumerate}
To meet these criteria, we propose the differential equation
\begin{equation}
\begin{split}
    \frac{dx_i}{dt} = r_i - x_i - a\sqrt{x} \cdot \exp\left(-bx_i\right) + w_\text{log} x_i \cdot \left( 1-x_i \right),
\label{eq:diffeq_ecolog_element}
\end{split}
\end{equation}
where $a$ and $b$ are parameters and $r_i$ is the limiting relative input as in the Leontief production function. The bifurcation diagram is given in Fig.~\ref{fig:eight}a. The first two terms in equation \ref{eq:diffeq_ecolog_element} represent a linear response to perturbations, similar to the Acclimate model (see the dotted line in Fig.~\ref{fig:eight}a). The third term is responsible for the nonlinear behaviour, causing tipping and hysteresis (see the dashed line in Fig.~\ref{fig:eight}a). However, an economic tipping element defined by these three terms would be inherently unstable. Even small perturbations would finally lead to a collapse of the node since perturbations are almost always growing due to the structure of the network. However, we know that the trade network is not that fragile. Therefore, we add a logistic growth term to the differential equation to stabilise the network on the individual node level with the weight $w_\text{log}$. Here, we argue that a certain flexibility in substituting inputs exists. Within limits, it is therefore possible to return to the original production due to the logistic growth term.
To illustrate an example, we choose $a = 4, b = 10$ for the parameters in Eq.~\ref{eq:diffeq_ecolog_element}. Therefore, the two bifurcation points lie at $r_1=0.4$ and $r_2=0.6$. The strength of the logistic growth term is chosen as $w_\text{log} = 0.2$ (see the blue line in Fig.~\ref{fig:eight}a).

In order to calculate the input term represented by $r_i$, every flow is normalised to the sum of flows from nodes of that sector to the target node. So the new weight $w_{c,s\rightarrow i}$ of a link from sector $s$ in country $c$ to node $i$ is given as
\begin{equation}
w_{c,s\rightarrow i} = \frac{f_{c,s\rightarrow i}} {\sum_{k} f_{k,s\rightarrow i}}. \label{eq:normalisation}
\end{equation}
With this, we can write the coupling term for the differential equation as
\begin{equation}
r_i = \min_{s \text{ in sectors}}\left\lbrace\sum_{c\text{ in countries}} w_{c,s\rightarrow i} \cdot x_{c,s}\right\rbrace.
\end{equation}

\begin{figure}
\centering
\resizebox{\columnwidth}{!}{\includegraphics{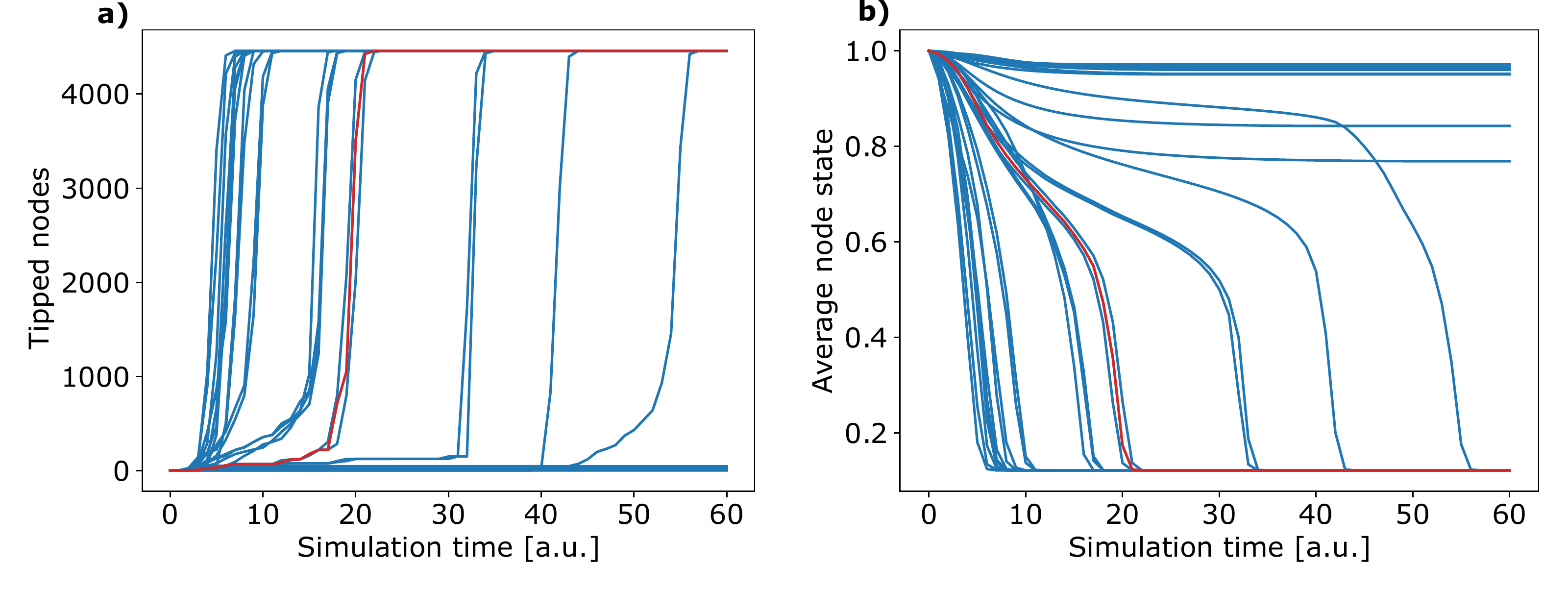}}
\caption{Tipping cascades with the economic tipping element defined by equation \ref{eq:diffeq_ecolog_element} on the normalised trade network. The cascade depicted in Fig.~\ref{fig:eight} is plotted in red. \textbf{a)} Number of tipped nodes over the course of the simulation time. Cascades either lead to tipping of almost all nodes (global tipping) or the dynamics of the tipping cascade stops growing after very few nodes are tipped. \textbf{b)} Average node state $\langle x\rangle$ over the course of the simulation time. At the onset of the tipping cascade, the average stage shows a sharp decline. Since a decline in the average state does not automatically mean that nodes were tipped, some average state time series stabilise while others show a global cascade.}
\label{fig:nine}
\end{figure}

To simulate cascades, we start with all nodes in the untipped state, here $x_i=1$ for all nodes $i$. We select a random starting node and tip it by setting its productivity to zero and then evolve the system with \emph{PyCascades}. We exemplify this for a cascade between three countries, where one node has been tipped artificially (see Fig.~\ref{fig:eight}b). The graph illustrates how the cascade propagates within and across the different countries forming densely connected network communities. Once the cascade reaches a country, most of that country's nodes tip almost at the same time. However, this gradual and sequential propagation of a tipping event is only one pattern of cascading behaviour observed. In Fig.~\ref{fig:nine}, we show cascades for 30 different start nodes, chosen such that a wide range of different tipping cascades can be observed. Fig.~\ref{fig:nine}a shows the number of tipped nodes, and Fig.~\ref{fig:nine}b the average node state $\langle x\rangle = \frac{1}{N}\sum_{i}^N x_i$.

\section{Conclusion}
\label{sec:conclusion}
In this work, we have outlined the software package \emph{PyCascades}, which is designed for simulating nonlinear dynamics, in particular tipping behaviour of interacting systems. For that purpose, two different types of tipping elements (Cusp and Hopf-bifurcation type models) are provided in \emph{PyCascades} as well as different paradigmatic complex network types (Erd\H{o}s-Rényi, Barabási-Albert, Watts-Strogatz networks) and a stochastic version of the tipping elements supplying Gaussian, Lévy and Cauchy noise. \emph{PyCascades} is written in the programming language Python and is written with an object-oriented architecture such that it remains flexible and can easily be adapted or extended to further applications or theoretical problems. 

However, a distinct limitation is, as of now, that only systems can be investigated, where the individual dynamics part can be separated from the interaction part. We also suspect that there is considerable potential for improvement in some technical details. For instance, more interaction types or multiplicative noise could be implemented. Another distinct constraint of \emph{PyCascades} is that only paradigmatic dynamics of tipping elements are implemented. In particular, it would be highly desirable to develop process-based tipping elements depending on the respective application.

All in all, due to modular setup, \emph{PyCascades} has the potential to contribute to relevant questions about the emergence of tipping cascades in various contexts, ranging from economics, ecology, climate science and beyond.

\newpage \clearpage
\section{Acknowledgements}
We thank Dorothea Kistinger for her work implementing the Hopf-bifurcation equation into \emph{PyCascades} and Benedikt Stumpf for many fruitful discussions during the development of \emph{PyCascades}. This work has been carried out within the framework of the IRTG 1740/TRP 2015/50-122-0 project funded by DFG and FAPESP. N.W. and R.W. acknowledge their financial support. This work is also part of PIK's FutureLab on Earth Resilience in the Anthropocene. N.W. is grateful for a scholarship from the Studienstiftung des deutschen Volkes. N.W., J.F.D., J.H. and R.W. are thankful for financial support by the Leibniz Association (project DominoES). A.S. and J.F.D. acknowledge support from the European Research Council Advanced Grant project ERA (Earth Resilience in the Anthropocene, ERC-2016-ADG-743080). J.F.D. is grateful for financial support by the Stordalen Foundation via the Planetary Boundary Research Network (PB.net) and the Earth League’s EarthDoc program, and S.W. acknowledges support by the German Federal Ministry of Education and Research (BMBF) under the research project CLIC (FKZ: 01LA1817C). The authors gratefully acknowledge the European Regional Development Fund (ERDF), the German Federal Ministry of Education and Research and the Land Brandenburg for supporting this project by providing resources on the high performance computer system at the Potsdam Institute for Climate Impact Research.

\textcolor{black}{\section{Author contribution statement}
N.W. designed the study together with R.W. and J.F.D.. N.W. performed the simulations and prepared the figures for this work with contributions from J.Kr. (in the section: Structure of the core of PyCascades) and V.W. (in the section: International Trade Network). N.W. led the writing of this work with inputs from all authors. J.Kr. developed the software package PyCascades with inputs and extensions from J.Ko., V.W. and N.W.. J.F.D. supervised this study.}

\section{Code and data availability}
The code of this work has been provided as the open source software package \emph{PyCas}-\emph{cades} written in the programming language Python. It is freely available on github under the doi: \href{http://dx.doi.org/10.5281/zenodo.4153102}{10.5281/zenodo.4153102}. To be used, \emph{PyCascades} is supplied with a 3-clause BSD license. In this repository, you can also find the code and data for the examples and applications that are discussed in this work. In case questions arise, please contact the corresponding authors of this study.

\bibliography{bibliography}


\newpage \clearpage

\end{document}